%% file: main.tex
\documentclass[aps,prl,twocolumn,showpacs,superscriptaddress,groupedaddress]{revtex4}  
\usepackage{graphicx}  
\usepackage{dcolumn}   
\usepackage{bm}        
\usepackage{amssymb}   
\usepackage{amsmath}
\usepackage{gensymb}

\usepackage[T1]{fontenc}
\usepackage{enumitem}
\usepackage{tikz}
\usepackage[caption=false]{subfig}
\usepackage{graphicx}

\showoutput
\begin{document}

\widetext
\leftline{Version : \today}

\hspace{5.2in}

\title{Inverse Saffman-Taylor experiments with particles lead to 
  capillarity driven fingering instabilities}
\input author_list.tex

\date{\today}

\begin{abstract}

Using air to displace a viscous fluid contained in Hele-Shaw cell can create a fingering pattern at the interface between the fluids, if the capillary number exceeds a critical value.  This Saffman-Taylor instability is revisited for the inverse case of a viscous fluid displacing air, when partially wettable hydrophilic particles are lying on the walls. Though the inverse case is otherwise stable, the presence of the particles results in a fingering instability at low capillary number. This capillary-driven instability is driven-by the integration of particles into the interface which results from the minimization of the interfacial energy. Both axisymmetric and rectangular geometries are considered in order to quantify this phenomenon.

\end{abstract}

\pacs{47,47.70.Qj,47.61.-k,47.61.Jd,47.55.N-,47.55.Kf}
\maketitle

The Saffman-Taylor instability \cite{Saffman,Lacalle,Levache} is a classical, interfacial instability that occurs when a low viscosity fluid displaces one of higher viscosity in a Hele-Shaw cell \cite{H-S}.  The instability generates a fingering phenomenon at the moving interface that leads to complex tree patterns.  Aside from the beauty of the structures, the attention devoted to this instability can be attributed to its widespread relevance in applications such as flows in porous media \cite{Sandnes,Sandnes1,Johnsen}, flame propagation \cite{Markstein,Pelce}, and growth of bacterial colonies \cite{Ben-Jacob}.  The instability results from the decrease of the flow resistance as the fluid of lower viscosity replaces the more viscous fluid; the inverse situation, wherein a highly viscous fluid is pushed into a cell filled with a weakly viscous fluid, is stable. Despite the large number of investigations of interfacial instabilities, the displacement of air by a viscous fluid in the presence of particles has not been reported in the literature.  Nevertheless, this case is of importance in cleaning processes when bacteria, spores, dust or particles are present on surfaces.

Here, we demonstrate the formation of fingering patterns emerging from the injection of a liquid into a cell filled with air in the presence of microparticles on the walls. This corresponds to the reverse situation compared to the common Saffman-Taylor instability. This instability relies on  the integration of partially wettable particles to the meniscus driven by the minimization of interfacial energy.  A reverse Saffman-Taylor instability in the presence of surfactants has been previously  reported in the literature by Chan {\it et al.} \cite{Chan} and then by Fernandez {\it et al.} \cite{Fernandez}, but, in contrast to these studies, the fingers observed in our studies with particles are not limited in depth.

Air-water interfaces can induce strong capillary forces able to remove micro-particles initially deposited on a wall surface. In nature, this phenomenon happens frequently when a raindrop falls on a lotus leaf. Due to the hydrophobicity of the surface and gravity, the water drop rolls off on the slanted leaf while also removing dust particles. This self-cleaning process is referred to as the ``Lotus-Effect'' \cite{Zhang,Barthlott} and is of primary interest to a wide variety of industrial processes including tube \cite{Farzam,Aramrak} or microelectronic silicon wafer cleaning \cite{Leenaars}, separation of minerals in the mining industry \cite{Min}, particle flotation \cite{Mason}, or even stabilization of bubbly liquids, foams and emulsions \cite{Binks,Ramsden,Binks1}. Once the micro-particles are removed from the surfaces, they are adsorbed and transported by the air-water interface. We show here how the presence of these particles can dramatically affect the stability of an interface.  

\begin{figure}[b!]
\includegraphics[width=0.4 \textwidth]{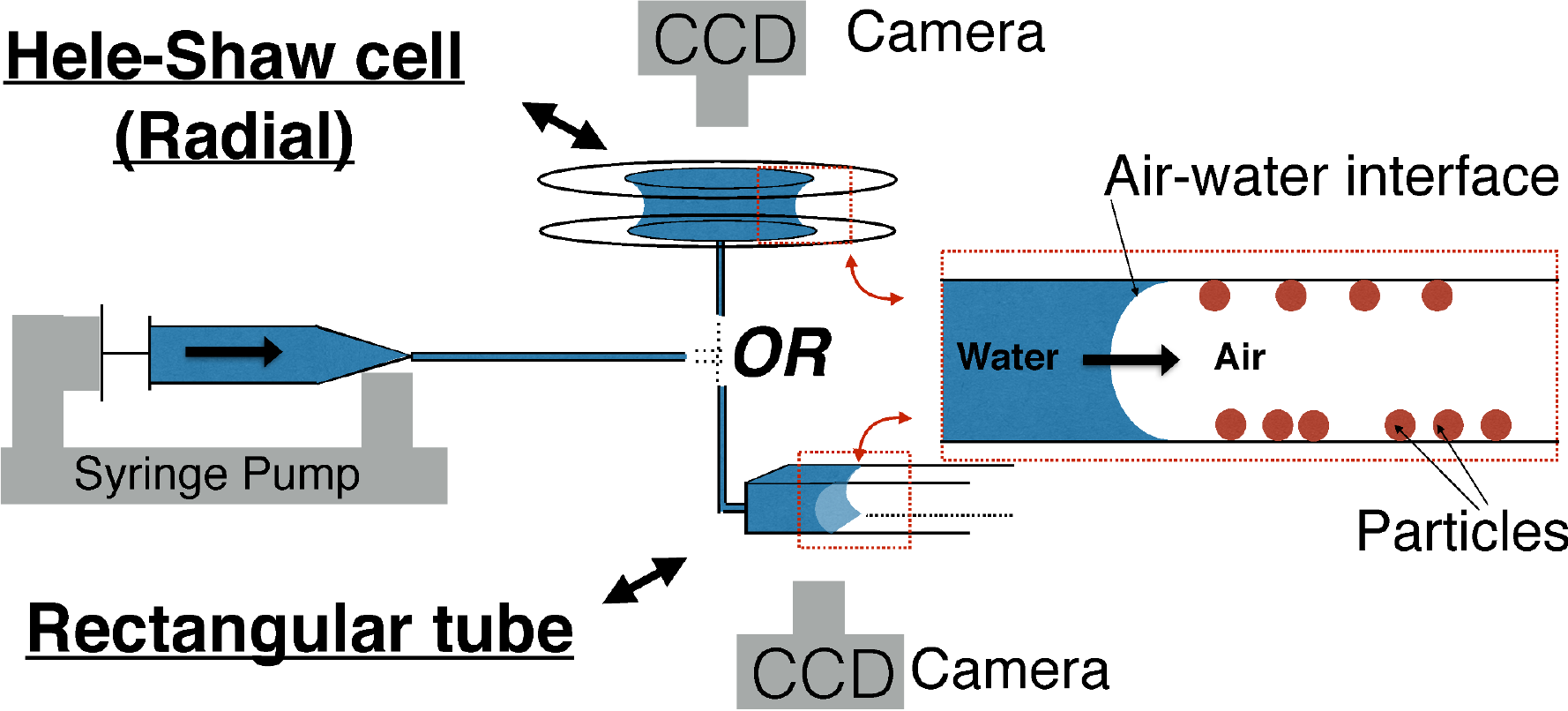} 
\captionsetup{justification=justified}
\caption[]{Schematic of the experimental setup: water is injected into air using a syringe pump either between two parallel plates (radial Hele-Shaw cell) or inside a rectangular tube. Micro-particles were deposited on the surfaces prior to water injection.}
\label{sketch}
\end{figure}

This ``particle driven'' fingering instability is first studied in a radial Hele-Shaw cell (see Fig.\ref{sketch}) which is contructed of two circular glass plates of radius 50 mm.  They are centered and placed on top of each other at a separation distance of $H$ that was varied between 0.1 and 1 mm. The plates were cleaned prior to each experiment with isopropanol, distilled water and acetone. Then, Rilsan (Polyamide 11) particles with an average radius of 15 $\mu$m were uniformly dispersed on both glass surfaces with a concentration $C$. This concentration corresponds to the ratio of the surface covered by the particles to the total surface and has been calculated by analyzing images of the plates obtained with an optical microscope. The particles have a density of $1.03$ g/cm$^3$ and are hydrophilic ($S_p<0, \theta_p=71 \degree \pm 3\degree$, with $S_p$ the spreading parameter and $\theta_p$ the static contact angle between the particles and deionized water \cite{Farzam}). Water is injected into this cell through a hole drilled at the center of the bottom plate at constant flow rates $Q$ ($2$, $20$, $200$ or $2000$ ml/h) and the evolution of the liquid/air interface was recorded with a CCD camera. For the experiments described above, the capillary number, Reynolds number, and Bond number are defined respectively as $Ca=\mu_lU/\gamma_{GL}$, $Re=\rho_lUH/\mu_l$ and $Bo = \rho_1gH^2/\gamma_{GL}$, where $\rho_l$, $\mu_l$ and $\gamma_{GL}$ are the density, viscosity and surface tension of the DI water.  Gravitational acceleration is $g$ and the values of $Bo$ range from approximately $10^{-3}$ to $10^{-1}$ depending on the plate separation. Both $Ca$ and $Re$ depend upon the velocity of the interface $U=Q/(2\pi r H)$, which is a function of the radius $r$.  At the inlet where $r=2$ mm, the maximum values are $Ca \approx 10^{-3}$ and $Re \approx 10^2$ and, at the outlet ($r=40$ mm), the minimum values are $Ca \approx 10^{-6}$ and $Re \approx 10^{-4}$.

\begin{figure}[b!]
\subfloat[\hfill]{
        \label{subfig:correct}
        \includegraphics[width=0.45 \textwidth]{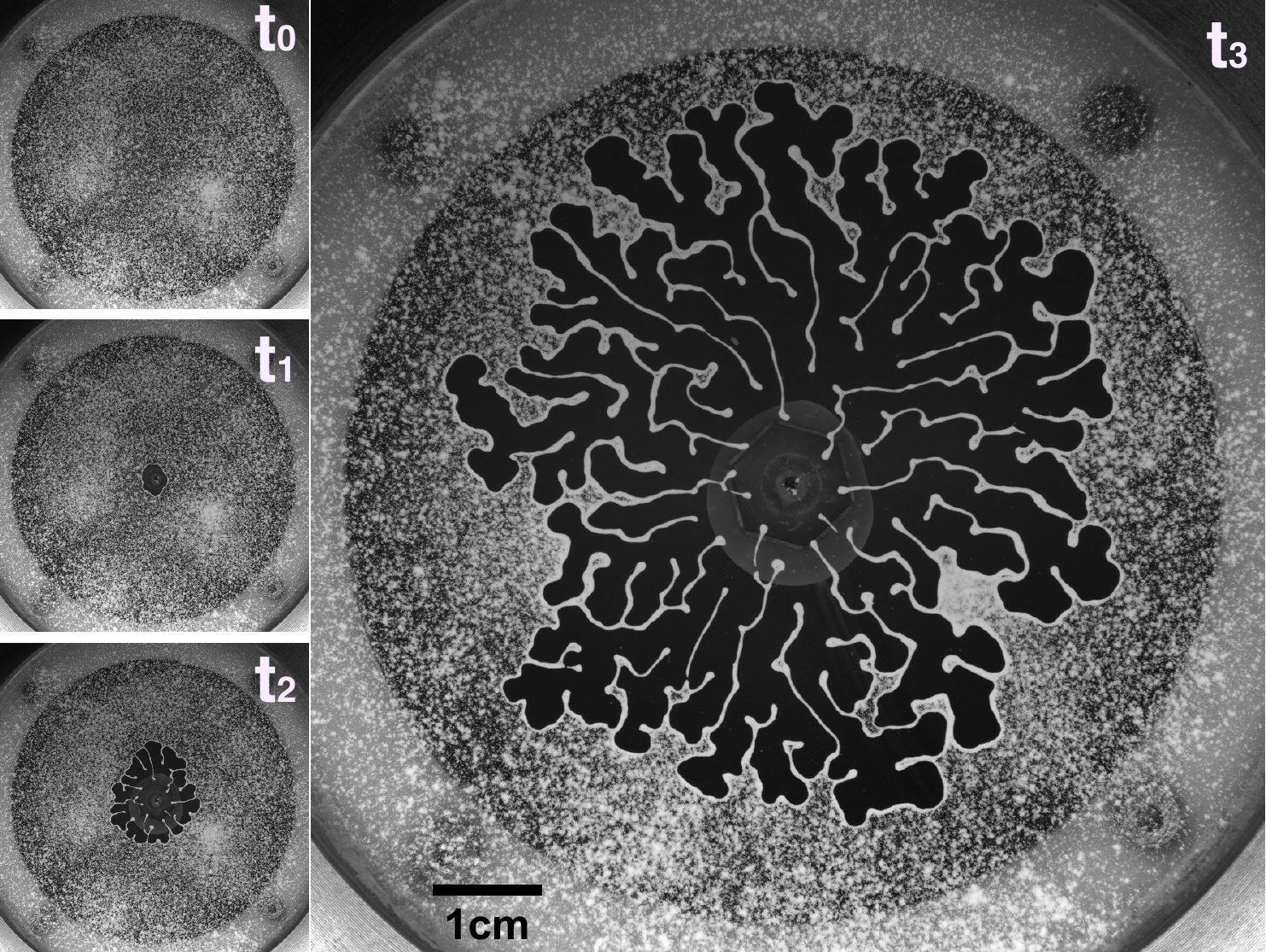} } 
\hfill
\subfloat[\hfill]{
        \label{subfig:notwhitelight}
        \includegraphics[width=0.45 \textwidth]{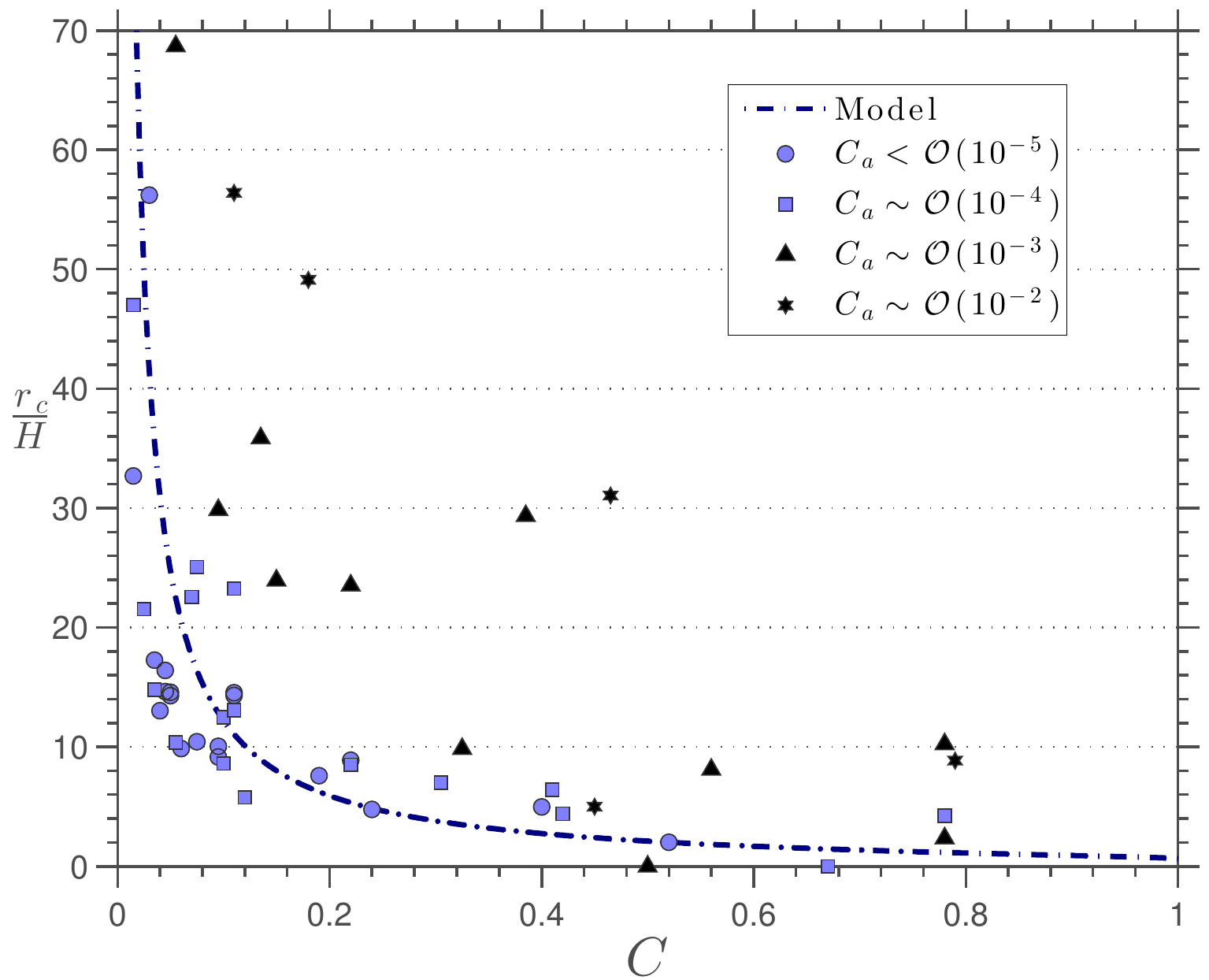} }    
\hfill
\captionsetup{justification=justified}
\caption[]{\textbf{(a)} Images showing a time sequence of the interfacial instability in a radial Hele-Shaw cell with a plate spacing of 150 $\mu$m, $C = 0.78$, flow rate $Q=200 \; ml/h$ ($Ca = 6.57*10^{-4}$) and $[t_0,t_1,t_2,t_3 = 0,3,7,31 s]$. \textbf{(b)} The critical radius at which the instabilities start in the radial Hele-Shaw cell.}   
\label{Experimentradial}
\end{figure}

As fluid is injected between the plates, the air-water interface starts to expand radially.  Microparticles encountered by the moving interface are captured and progressively cover the meniscus. When a critical radius $r_c$ is reached, destabilization of the interface occurs and the formation of fingering patterns begins (see Fig. \ref{Experimentradial}.a).  To better understand this phenomenon, the critical radius was measured for different particle concentrations $C$, gap widths $H$ and flow rates (quantified by the capillary number $Ca$ at the critical radius $r_c$); the data is shown in Fig. \ref{Experimentradial}.b.

\captionsetup{singlelinecheck = off}

\begin{figure*}[tbp]

\includegraphics[width=1 \textwidth]{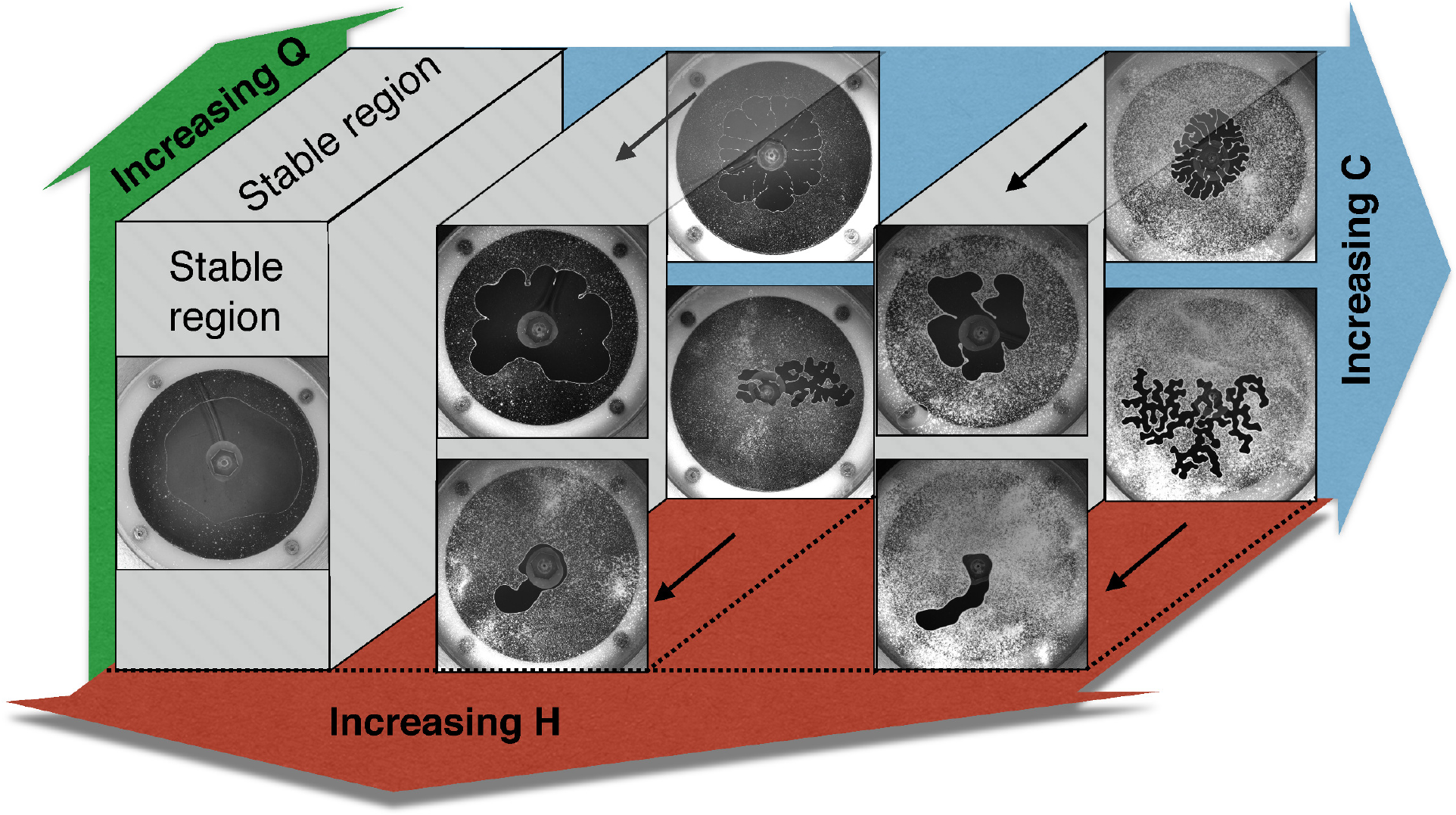} 
\captionsetup{justification=justified}
\caption[]{Diagram of the different instability patterns as a function of the concentration of the particles $C$, the flow rate $Q$ and the gap between the plates $H$.}
\label{Patterns}

\end{figure*}

In the present case, the interface instability seems to occur once the meniscus is entirely covered with particles.  The critical radius at which this occurs can be estimated from the following formula,
\begingroup\makeatletter\def\f@size{8.7}\check@mathfonts
\begin{equation*}
r_c / H = \frac{(\phi \pi -2C) + \sqrt{(\phi \pi-2C)^2 + 3C\phi(\pi -2) -3C^2}}{4C},
\end{equation*}
\endgroup
where $\phi$ is the specific surface area of the interface as defined by Torquato \cite{S. Torquato} ($\phi \simeq 0.8$ for a random organisation of the particles on the interface).  This prediction results from equating the surface area of the liquid interface, assuming a contact angle of $0\degree$ between the water and the plates, and the surface area of the particles that have been surpassed by the interface as it expands purely radially (i.e\. prior to the instability).  Figure \ref{Experimentradial}.b shows that the calculation of $r_c$ closely matches the experimental results at low capillary numbers ($Ca < 10^{-4}$), when interfacial phenomena largely dominates the dynamical phenomena, supporting the hypothesis.  Indeed, it is well-known \cite{Mason,Flury,Farzam} that hydrophilic particles are naturally collected by a liquid/air interface to minimize the interfacial energy. However, once the expanding interface becomes entirely covered by particles,  integrating additional particles into the interface is not possible while also maintaining a purely radial growth.  Instead, either the air liquid-interface will start expanding in a different way to incorporate new particles or the particles will remain in the gas or enter the liquid phase. The former case is always observed in the present experiments. 

For concentrations $C \le 0.8$, the integration of particles is achieved through the development of a fingering pattern that increases the interface to volume ratio as compared to the radial configuration. At very high particle concentrations ($C \ge 0.85$), the intergration of particles is generally achieved through the deposition of a liquid film ahead of the meniscus, similar to what was observed previously in cylindrical tubes \cite{Farzam} and will not be further discussed here. For the higher capillary numbers, the measurement of a higher critical radius than predicted from the above calculation is indicative of the role played by the fluid flow in the formation of the fingering patterns.

We further investigated the structure of the patterns (finger widths and numbers, symmetry) as a function of the experimental parameters $C$, $H$ and $Q$ (see Fig. \ref{Patterns}). The width and number of fingers mainly depend on the plate separation distance $H$ and the particle concentration $C$, while the symmetry of the patterns is mainly determined by the flow rate $Q$. For $Q \gtrsim 20ml/h \equiv Ca \gtrsim 10^{-4}$, symmetric radial fingers are observed, similar to the viscous fingers produced by Saffman-Taylor instabilities. For lower capillary numbers, fingers are formed in a prefential direction similar to labyrinth patterns produced by the drainage of a granular-fluid system in two dimensional confinement \cite{Sandnes}.

To better determine the width of the fingers $L_D$ produced by this instability as a function of the particle concentration, we performed experiments in a rectangular Borosilicate capillary tube of height $H=0.40$ mm and width $W=10 H$.  The inner walls were first degreased by sonication while suspended in acetone, then isopropanol, and finally in dichloromethane for five minutes each. Next, the tubes were dried under nitrogen flow and submerged in a freshly prepared piranha solution (sulfuric acid $H_2SO_4$ + hydrogen peroxide $H_20_2$) at 100$\degree$C for one hour. The channels were rinsed thoroughly with deionized water and dried in an oven at 120$\degree$C for one hour.  Then, the inner walls were covered by Rilsan (Polyamide11) particles by gently blowing them into the channel with an air jet.  Water was injected through the tube at a constant flow rate $Q=0.1$ ml/h,  corresponding to a capillary number $Ca=\mu_lQ/(\gamma_{GL}WH)=2.38 \times 10^{-7}$, while recording the dynamics of the moving meniscus as described previously.

\begin{figure}[t!]

\subfloat[\hfill]{
        \label{subfig:notwhitelight}
        \includegraphics[width=0.46 \textwidth]{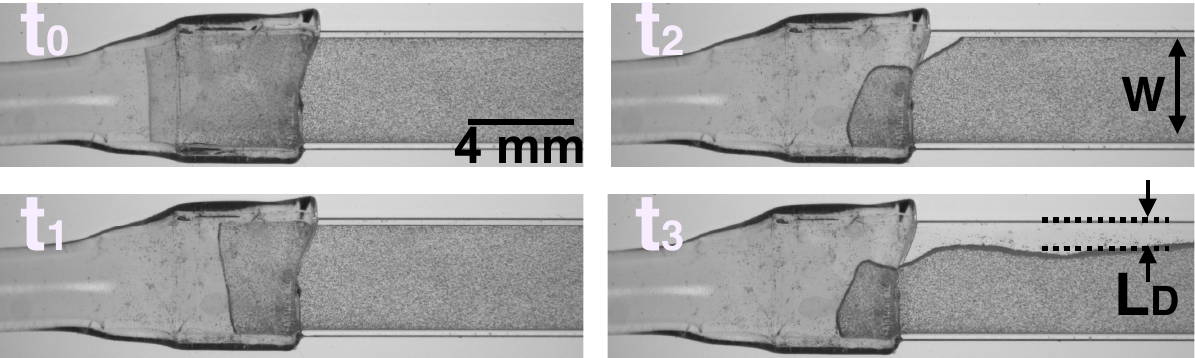} }    
\hfill
\subfloat[\hfill]{
        \label{subfig:correct}
        \includegraphics[width=0.46 \textwidth]{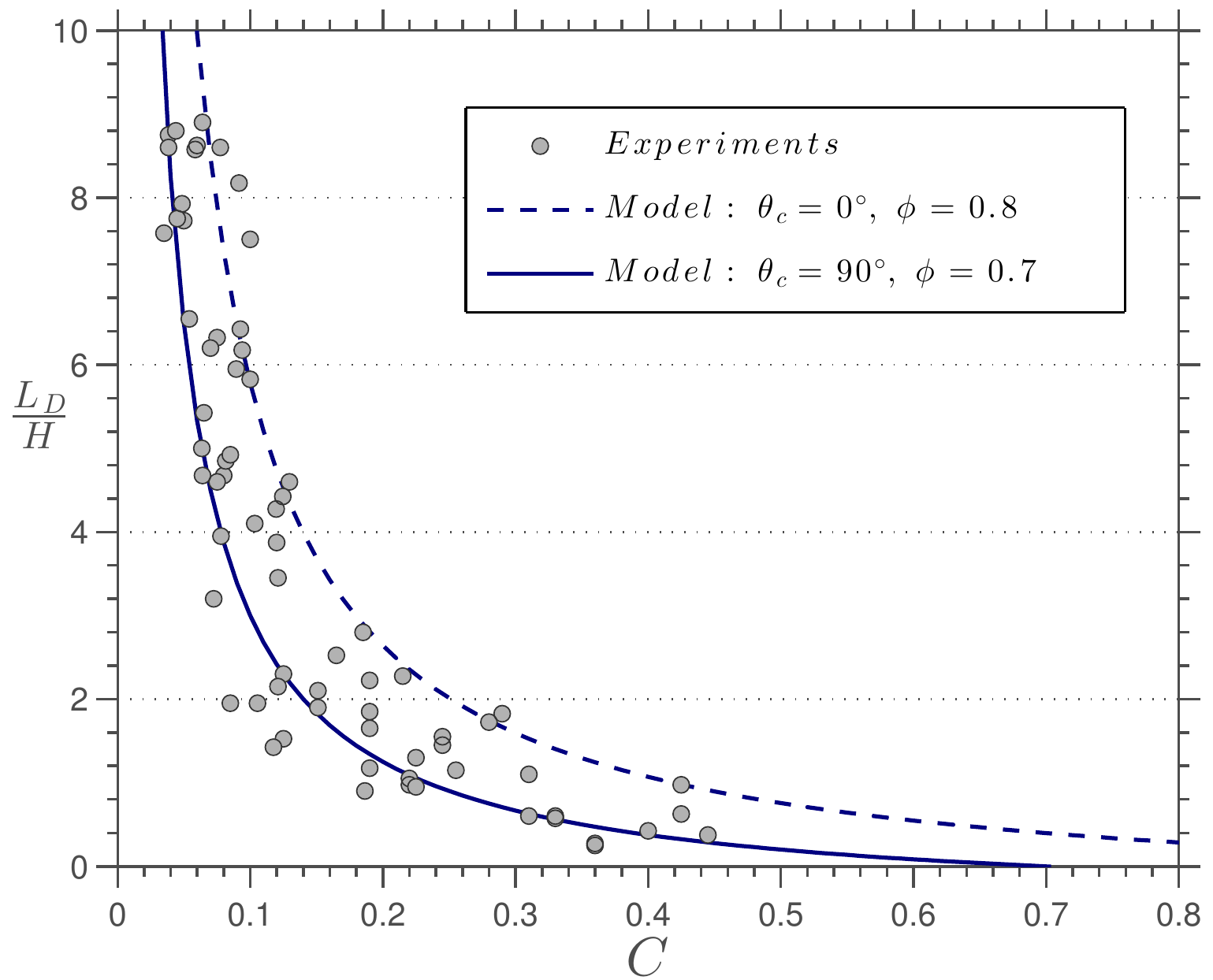} } 
\hfill
\subfloat[\hfill]{
        \label{subfig:notwhitelight}
        \includegraphics[width=0.46 \textwidth]{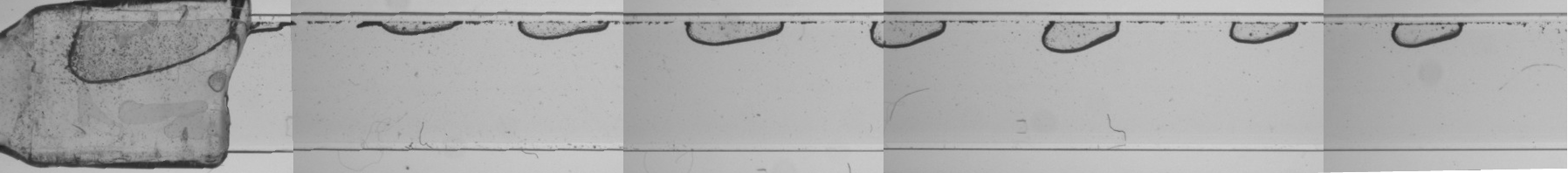} }    
\hfill
\captionsetup{justification=justified}

\caption[]{\textbf{(a)} A time sequence of images showing the interface instability in a rectangular capillary tube with a particle concentration of $C = 0.3$ and capillary number $Ca=2.38 \times 10^{-7}$. The air appears dark gray and the water is light gray. \textbf{(b)} Measured and predicted finger width as a function of the concentration of the particles in the rectangular tube. \textbf{(c)} Semi-armoured bubbles form when the finger width exceeds the width of the channel: $L_D/W > 1$.}
\label{results2}
\end{figure}

As water is injected into the tube, particles initally are collected by the meniscus at the relatively short times of $t_o$ and $t_1$ as seen in Fig.\ \ref{results2}.a.  For longer times $t_2$ and $t_3$, an instability occurs and leads to the formation of a liquid finger as also observed in the radial Hele-Shaw geometry.  Unlike the latter geometry however, the fingers of water in the rectangular channel retain a uniform width $L_D$ at constant concentration $C$.  The measurements of $L_D$, as given in Fig.\ \ref{results2}.b, indicate an inverse relationship with the concentration $C$.  It is interesting to note that when the finger width $L_D$ exceeds the width $W$, the meniscus touches the opposite wall and semi-armoured bubbles, attached to the walls, are formed periodically as seen in Fig.\ \ref{results2}.c.

As before, we suppose that the interface becomes unstable once covered by the particles and that the subsequent formation of fingers is driven by interfacial energy considerations. A theoretical liquid finger width, which corresponds to the limit size allowing all particles encountered by the meniscus to be captured, can be derived: $ L_D= H \phi \pi / 4C$ for a liquid-air-wall contact angle of $\theta_c =0\degree$ and $ L_D= H\phi/2C -H/2$ for a liquid-air-wall contact angle of $\theta_c =90\degree$. Comparisons of the measurements with the predictions for $L_D$ are shown in Fig.\ \ref{results2}.b as a function of the particle concentration $C$ for different model parameters $\theta_c$ and $\phi$, as these two parameters likely vary depending on the surface cleanness and the  distribution of the particles on the meniscus. The model captures the qualitative trends of the experimental data and the data is accurately bounded by the predictions of the limiting case of $\theta_c=0\degree$ with $\phi=0.8$, providing support for the idea that the finger width is simply fixed by an interfacial energy minimization process.

To conclude, we have demonstrated that the presence of partially wettable particles on the walls can dramatically affect the dynamics of a liquid finger that is injected into a Hele-Shaw cell filled with air.  Generally, a fingering pattern forms, even though the displacement of air with a viscous fluid is typically stable, unlike the classical Saffman-Taylor instability. The destabilization of the interface occurs due to interfacial energy minimization, which requires that all particles intersected by the meniscus are collected. As a consequence, the critical radius at which the instability occurs and the width of the fingers can be calculated by simply balancing the space available on the meniscus with the area needed to accomodate additional particles encountered by the liquid-air interface during its motion.

We gratefully acknowledge the support from Marie Curie International Research Staff Exchange Scheme (IRSES) Fellowship project titled Patterns and Surfaces (No. 269207) within the 7th European Community Framework Programme.

\input{bibliographie.tex}
\end{document}

%% file: author_list.tex
\affiliation{Univ. Lille, CNRS, ECLille, ISEN, Univ. Valenciennes, UMR 8520 - IEMN, F-59000 Lille, France}
\affiliation{Department of Chemical Engineering, University of Florida, Gainesville, Florida ,USA}
\affiliation{INRA, UR638, Villeneuve d'Ascq, France}
\author{Ilyesse Bihi} \affiliation{Univ. Lille, CNRS, ECLille, ISEN, Univ. Valenciennes, UMR 8520 - IEMN, F-59000 Lille, France} \affiliation{Department of Chemical Engineering, University of Florida, Gainesville, Florida ,USA}
\author{Michael Baudoin} \email[Electronic address: ]{ michael.baudoin@univ-lille1.fr} \affiliation{Univ. Lille, CNRS, ECLille, ISEN, Univ. Valenciennes, UMR 8520 - IEMN, F-59000 Lille, France}
\author{Jason.E Butler} \affiliation{Department of Chemical Engineering, University of Florida, Gainesville, Florida ,USA}
\author{Christine Faille} \affiliation{INRA, UR638, Villeneuve d'Ascq, France}
\author{Farzam Zoueshtiagh} \email[Electronic address: ]{ farzam.zoueshtiagh@univ-lille1.fr}\affiliation{Univ. Lille, CNRS, ECLille, ISEN, Univ. Valenciennes, UMR 8520 - IEMN, F-59000 Lille, France}

\vskip 0.25cm